\documentclass[epj]{webofc}%
\usepackage[varg]{txfonts}
\usepackage{amsmath}
\usepackage{amsfonts}
\usepackage{amssymb}
\usepackage{graphicx}%
\providecommand{\U}[1]{\protect\rule{.1in}{.1in}}
\wocname{EPJ Web of Conferences}
\woctitle{ICNFP 2017}

\begin{document}

\title{Toward a QFT treatment of nonexponential decay}
\author{Francesco Giacosa\inst{1,2}\fnsep\thanks{\email{fgiacosa@ujk.edu.pll}} }

\selectlanguage{english}

\institute{Institute of Physics, Jan Kochanowski University, 25-406 Kielce, Poland
\and
Institute for Theoretical Physics, Johann Wolfgang Goethe University, 60438 Frankfurt am Main, Germany}

\abstract{ We study  the properties of the survival probability of an unstable quantum state described by a
Lee Hamiltonian. This theoretical approach resembles closely Quantum Field Theory (QFT): one can introduce
in a rather simple framework the concept of propagator and Feynman rules, Within this context, we re-derive
(in a detailed and didactical way) the well-known result according to which the
amplitude of the survival probability  is the Fourier transform of the energy distribution (or spectral function)
of the unstable state (in turn, the
energy distribution is proportional to the imaginary part of the propagator of the unstable state).
Typically, the survival probability amplitude is
the starting point of many studies of non-exponential decays. This work represents a further step toward the evaluation of the
survival probability amplitude in genuine relativistic QFT.  However, although many similarities exist, QFT
presents some differences w.r.t. the Lee Hamiltonian which should be studied in the future.
}
\maketitle
\section{Introduction}

\label{intro}

Quantum decays are a common phenomenon in particle, nuclear, and atomic
physics \cite{ghirardi,fprev,pdg}. A typical starting point for the discussion
of the decay of the amplitude for the survival probability of a certain
unstable state $S$,%

\begin{equation}
a_{S}(t)=\int_{m_{th}}^{+\infty}\mathrm{dm}d_{S}(m)e^{-imt}\text{ ,}\label{as}%
\end{equation}
where $d_{S}(m)$ is the so-called energy (or mass) distribution ($\mathrm{dm}%
d_{S}(m)$ is the probability that the unstable state has an energy (or mass)
between $m$ and $m+dm$ and $m_{th}$ the lowest energy of the system). Under
general assumptions, one can show that the survival probability $p_{S}%
(t)=\left\vert a_{S}(t)\right\vert ^{2}$ is not exponential both at short
times (where $p_{S}^{\prime}(t\rightarrow0)=0$) and at long times (where a
power low is realized). For these deviations to occur, it is enough that a
low-energy threshold is present and that $d_{S}(m)$ is not of the Breit-Wigner
type \cite{vw}, see also Refs.
\cite{ghirardi,fprev,khalfin,facchiprl,winter,volya,urbanolast} and refs.
therein. The short-time behavior leads to the so-called Quantum\ Zeno Effect
(QZE): multiple collapse measurements freeze the time evolution, thus
preventing the decay to take place \cite{dega,misra,koshinorev,gppra}. Note,
very often $p_{S}(t)$ is expressed as $1-t^{2}/\tau_{Z}^{2}+...$($\tau_{Z}$ is
the Zeno time), but the weaker requirement of a zero derivative of $p_{S}(t)$
at $t=0$ is sufficient. Experimentally, deviations from the exponential law
have been measured at short times in Ref. \cite{raizen} (for the corresponding
QZE see Ref. \cite{raizen2}) and at long times in\ Ref. \cite{rothe} (for an
indirect proof through data on beryllium decays, see Ref. \cite{kelkar}).

Lee Hamiltonians \cite{lee} (LH) represent a useful theoretical framework for
the study of decays, e.g. Refs.
\cite{fprev,facchiprl,vanhove,duecan,giacosapra} and refs. therein. This
approach resembles very closely Quantum\ Field Theory (QFT). Similar
Hamiltonians have been used in various areas of physics, which go from atomic
physics and quantum optic \cite{vw,ford,jc} to QCD \cite{liu}.

The issue of non-exponential decay in a pure QFT framework is still debated:
while in\ Ref. \cite{maiani} a negative result was found (also in the case of
super-renormalizable Lagrangian), in\ Ref. \cite{duecan,zenoqft} a different
result is obtained: it is argued that also in QFT Eq. (1) holds and short- and
long-time deviations take place.

While the final goal is the derivation of Eq. (1), and hence of
non-exponential decay, in a genuine QFT relativistic environment, in this work
we take a more humble intent. We aim to recall in a detailed (and also
didactical) way how Eq. (1) emerges when using Lee Hamiltonians. In
particular, we shall also show that $d_{S}(m)$ is the mass distribution of the
decaying particle. Moreover, we establish a link between a discrete and
continuous base of final states and between the basis of the unperturbed and
full Hamiltonians. This study is intended to be useful for further analyses on
non-exponential decays.

The article is organized as follows: in\ Sec. 2 we present the Lee
Hamiltonian, both in the discrete and in the continuous cases. Then, in Sec. 3
we study the time evolution of an unstable state: the amplitude of the
survival probability is expressed first as the Fourier transformation of the
propagator and then of the energy distribution. Finally, in Sec. 4 we present
conclusions and outlook.

\section{The Lee Hamiltonian's approach}

We present here the Lee Hamiltonian (LH), first using an infinite discrete set
of decay products and then performing the limit to the continuous case.

\subsection{Discrete LH}

Let us consider the quantum state $\left\vert S\right\rangle $ as the unstable
state that we aim to investigate. In particular, we study its time evolution
after its preparation at $t=0.$ The state $\left\vert S\right\rangle $
interacts with an `infinity' of other states, denoted as:
\begin{equation}
\left\vert k_{n}\right\rangle \text{ with }k_{n}=\frac{2n\pi}{L}\text{ and
}n=0,\pm1,\pm2,...\text{ ,}%
\end{equation}
where the quantity $L$ (with the dimension of energy$^{-1}$) can be thought as
the length of the linear box in which we place our system. The physical
results should not depend on $L,$ if it is large enough. The quantities
$k_{n}$ `look like' momenta, see below. Finally, the basis of the Hilbert
space of our quantum problem reads:%
\begin{equation}
\text{Basis of the Hilbert space }\mathcal{H}\text{: }\left\{  \left\vert
S\right\rangle ,\left\vert k_{0}\right\rangle ,\left\vert k_{1}\right\rangle
,\left\vert k_{-1}\right\rangle ,...\right\}  \equiv\left\{  \left\vert
S\right\rangle ,\left\vert k_{n}\right\rangle \right\}
\end{equation}
with the usual orthonormal and completeness relations:
\begin{equation}
\left\langle S|S\right\rangle =1\text{ , }\left\langle S|k_{n}\right\rangle
=0\text{ },\text{ }\left\langle k_{n}|k_{m}\right\rangle =\delta_{nm}\text{ ;
}\left\vert S\right\rangle \left\langle S\right\vert +\sum_{n}\left\vert
k_{n}\right\rangle \left\langle k_{n}\right\vert =1_{\mathcal{H}}\text{ .}%
\end{equation}

The Lee Hamiltonian of the system consists of two pieces:
\begin{equation}
H=H_{0}+H_{1}%
\end{equation}
where $H_{0}$ describes the free (non-interacting) part while $H_{1}$ mixes
$\left\vert S\right\rangle $ with all $\left\vert k_{n}\right\rangle $
\begin{equation}
H_{0}=M_{0}\left\vert S\right\rangle \left\langle S\right\vert +\sum
_{n=0,\pm1,...}\omega(k_{n})\left\vert k_{n}\right\rangle \left\langle
k_{n}\right\vert \text{ ; }H_{1}=\sum_{n=0,\pm1,...}\frac{gf(k_{n})}{\sqrt{L}%
}\left(  \left\vert S\right\rangle \left\langle k_{n}\right\vert +\left\vert
k_{n}\right\rangle \left\langle S\right\vert \right)  \text{ .}%
\end{equation}

Following comments are in order:

\begin{itemize}
\item The quantities $M_{0},$ $\omega(k_{n}),$ $gf(k_{n})$ are real.

\item The Hamiltonian $H$ is Hermitian.

\item Dimensions: $M_{0}$ and $\omega(k_{n})$ have dimensions [energy], while
$g$ has dimensions [energy$^{+1/2}$]

\item The energy $M_{0}$ is the bare energy of the level $\left\vert
S\right\rangle $. In particle physics, it is the bare mass at rest.

\item The energy $\omega(k_{n})$ is the bare energy of the state $\left\vert
k_{n}\right\rangle ,$ see below.

\item The coupling constant $g$ measures the strength of the interaction; the
form factor $f(k_{n})$ modulates the interaction. In practice, each mixing
$\left\vert S\right\rangle \longleftrightarrow\left\vert k_{n}\right\rangle $
has its own coupling constant $gf(k_{n}).$

\item The factor$\sqrt{L}$ is introduced for future convenience: it is
necessary for a smooth continuous limit $L\rightarrow\infty$.

\item For notational simplicity, $\sum_{n=0,\pm1,...}$can be also expressed
simply as $\sum_{n}$.
\end{itemize}

Further discussion concerns the interpretation and the energy $\omega(k_{n})$.

\textbf{Interpretation:} The state $\left\vert S\right\rangle $ represents an
unstable particle $S$ in its rest frame and the state $\left\vert
k_{n}\right\rangle $ represents a possible final state of the decay of $S.$ In
the simplest case of a two-body decay, the state $\left\vert k_{n}%
\right\rangle $ represents \textbf{two} particles emitted by $S$ flying
back-to-back:%
\begin{equation}
S\rightarrow\varphi_{1}+\varphi_{2}\text{ .}%
\end{equation}
In the case of one spacial dimension, $k_{n}$ can be interpreted as the
momentum of $\varphi_{1}$, while $-k_{n}$ is the momentum of $\varphi_{2}$.
Schematically: $\left\vert k_{n}\right\rangle \equiv\left\vert \varphi
_{1}(k_{n}),\varphi_{2}(-k_{n})\right\rangle $. In this way, the total
momentum of $\left\vert k_{n}\right\rangle $ is still zero, as it must. (The
3D extension is straightforward). As examples of such a process, we may think
of: (i) The neutral pion $\pi^{0}$ decays into two photons: $\pi
^{0}\rightarrow\gamma\gamma.$ Then, $\pi^{0}$ in its rest frame corresponds to
$\left\vert S\right\rangle ,$ while $\gamma\gamma$ corresponds to $\left\vert
k_{n}\right\rangle $ (one photon has momentum $k_{n},$ the other $-k_{n}$).
(Note, a very large number of two-body decays is listed in the PDG
\cite{pdg}). (ii) An excited atom $A^{\ast}$ decays into the-ground state atom
$A$ emitting a photon $\gamma$: $A^{\ast}\rightarrow A\gamma.$ In this case,
$A^{\ast}$ is the sate $\left\vert S\right\rangle ,$ while $\left\vert
k_{n}\right\rangle $ represents the joint system of the ground-state atom $A$
and the photon.

\textbf{Function }$\omega(k_{n})$: as mentioned above, the function
$\omega(k_{n})$ represents the energy of the state $\left\vert k_{n}%
\right\rangle $. In the case of a two-body decay its form is given by
\begin{equation}
\omega(k_{n})=\sqrt{k_{n}^{2}+m_{1}^{2}}+\sqrt{k_{n}^{2}+m_{2}^{2}}\text{ ,}%
\end{equation}
where $m_{1}$ is the mass of $\varphi_{1}$ and $m_{2}$ of $\varphi_{2}$.
Clearly, $\omega(k_{n})\geq m_{1}+m_{2}=m_{th}$, where $m_{th}$ represents the
lowest energy of the $\left\vert k\right\rangle $ states. In the two-photon
decay such as the process (i) described above, one has $m_{1}=m_{2}=0,$ hence
$\omega(k_{n})=2\left\vert k_{n}\right\vert \geq0=m_{th}$. In an atomic decay
of the type $A^{\ast}\rightarrow A+\gamma$, one has $m_{1}=0,$ and
$m_{2}=M_{A},$ hence:%
\begin{equation}
\omega(k_{n})\simeq\left\vert k_{n}\right\vert +M_{A}.
\end{equation}
In this case, one could also subtract a constant term, $H_{0}\rightarrow
H_{0}-M_{A}1_{\mathcal{H}},$ out of which $\omega(k_{n})\simeq\left\vert
k_{n}\right\vert $.

\subsection{Continuous LH}

The limit $L\rightarrow\infty$ implies that the variable $k_{n}$ becomes
continuos:
\begin{equation}
k_{n}=\frac{2\pi n}{L}\rightarrow k\subset(-\infty,+\infty).
\end{equation}
As usual, when $L$ is sent to infinity sums turn into integrals:%
\begin{equation}
\sum_{n}=\frac{L}{2\pi}\sum_{n}\frac{2\pi}{L}\rightarrow\frac{L}{2\pi}%
\int_{-\infty}^{+\infty}\mathrm{dk}=L\int_{-\infty}^{+\infty}\frac
{\mathrm{dk}}{2\pi}\text{ ,}%
\end{equation}
where $\delta k=2\pi/L$ has been introduced in order to generate the
differential $dk$. Next, we turn to the kets $\left\vert k\right\rangle $ in
the continuous limit, for which we expect that $\left\langle k_{1}%
|k_{2}\right\rangle =\delta(k_{1}-k_{2})$. To this end, let us write down the
following $L$-dependent discrete representation of the Dirac-delta function:%

\begin{equation}
\delta_{L}(k_{n})=\int_{-L/2}^{L/2}\frac{\mathrm{dx}}{2\pi}e^{ik_{n}%
x}=\left\{
\begin{array}
[c]{c}%
0\text{ for }n\neq0\\
\frac{L}{2\pi}\text{ for }n=0
\end{array}
\right.  \text{ .}%
\end{equation}
In the limit $L\rightarrow\infty$ one obtains (for an arbitrary function
$u(k)$):
\begin{equation}
u(0)=\sum_{n}\delta k\delta_{L}(k_{n})u(k_{n})\rightarrow\int_{-\infty
}^{+\infty}\mathrm{dk}\delta(k)u(k)=u(0)
\end{equation}
showing that $\delta(k)=\lim_{L\rightarrow\infty}\delta_{L}(k_{n})$ holds.
Finally, the quite subtle link between $\left\vert k_{n}\right\rangle $ and
$\left\vert k\right\rangle $ is given by:%
\begin{equation}
\left\vert k_{n}\right\rangle \overset{L\rightarrow\infty}{=}\sqrt{\frac{2\pi
}{L}}\left\vert k\right\rangle \text{ .}%
\end{equation}
Namely:
\begin{equation}
\left\langle k_{1}|k_{2}\right\rangle =\lim_{L\rightarrow\infty}\frac{L}{2\pi
}\left\langle k_{n_{1}}|k_{n_{2}}\right\rangle =\lim_{L\rightarrow\infty
}\left\{
\begin{array}
[c]{c}%
0\text{ for }n_{1}\neq n_{2}\\
\frac{L}{2\pi}=\delta_{L}(0)\text{ for }n_{1}=n_{2}%
\end{array}
\right.  =\delta(k_{1}-k_{2})\text{ ,}%
\end{equation}
as desired. (Note, in 3D we have $\sum_{\mathbf{k}}\rightarrow V\int
\frac{d^{3}k}{(2\pi)^{3}}$ ,where $V=L^{3},$ and $\left\vert \mathbf{k}%
=2\pi\mathbf{n}/L\right\rangle \rightarrow(2\pi)^{3/2}/\sqrt{V}\left\vert
\mathbf{k}\right\rangle $). It is also quite peculiar that the dimension of
the ket changes when considering the limit $L\rightarrow\infty$:%
\begin{equation}
\dim[\left\vert k_{n}\right\rangle ]=[\text{Energy}^{0}]\text{ (dimensionless)
, }\dim[\left\vert k\right\rangle ]=[\text{Energy}^{-1/2}]\text{ .}%
\end{equation}
Then, the continuos Hilbert space is given by $\mathcal{H}=\left\{  \left\vert
S\right\rangle ,\left\vert k\right\rangle \right\}  $ with%
\begin{equation}
\left\langle S|S\right\rangle =1\text{ , }\left\langle S|k\right\rangle
=0\text{ },\text{ }\left\langle k_{1}|k_{2}\right\rangle =\delta(k_{1}%
-k_{2})\text{ .}%
\end{equation}
We also check the completeness relation:
\begin{equation}
1_{\mathcal{H}}=\left\vert S\right\rangle \left\langle S\right\vert +\sum
_{n}\left\vert k_{n}\right\rangle \left\langle k_{n}\right\vert =\left\vert
S\right\rangle \left\langle S\right\vert +\sum_{n}\delta k\left(  \sqrt
{\frac{L}{2\pi}}\left\vert k_{n}\right\rangle \left\langle k_{n}\right\vert
\sqrt{\frac{L}{2\pi}}\right)  \overset{L\rightarrow\infty}{\rightarrow
}\left\vert S\right\rangle \left\langle S\right\vert +\int_{-\infty}^{+\infty
}dk\left\vert k\right\rangle \left\langle k\right\vert =1_{\mathcal{H}}%
\end{equation}
Finally, we are ready to present the Lee Hamiltonian $H=H_{0}+H_{1}$ in the
continuous limit:%

\begin{equation}
H_{0}=M\left\vert S\right\rangle \left\langle S\right\vert +\int_{-\infty
}^{+\infty}\mathrm{dk}\omega(k)\left\vert k\right\rangle \left\langle
k\right\vert \text{ , }H_{1}=\int_{-\infty}^{+\infty}\mathrm{dk}\frac
{gf(k)}{\sqrt{2\pi}}\left(  \left\vert S\right\rangle \left\langle
k\right\vert +\left\vert k\right\rangle \left\langle S\right\vert \right)
\text{ .}\nonumber
\end{equation}
One can verify that the dimensions is preserved. For instance:
\begin{equation}
\dim\left[  dk\omega(k)\left\vert k\right\rangle \left\langle k\right\vert
\right]  =\dim[dk]\dim[\omega(k)]\dim^{2}[\left\vert k\right\rangle
]=[\text{Energy}][\text{Energy}][\text{Energy}^{-1}]=[\text{Energy}]\text{ .}%
\end{equation}

\section{Determination of the survival probability}

\subsection{Time evolution operator}

The Schr\"{o}dinger equation (in natural units)%
\begin{equation}
i\frac{\partial\left\vert \psi(t)\right\rangle }{\partial t}=H\left\vert
\psi(t)\right\rangle
\end{equation}
can be univocally solved for a certain given initial state
\begin{equation}
\left\vert \psi(0)\right\rangle =c_{S}\left\vert S\right\rangle +\sum_{n}%
c_{n}\left\vert k_{n}\right\rangle \overset{L\rightarrow\infty}{\equiv}%
c_{S}\left\vert S\right\rangle +\int_{-\infty}^{+\infty}\mathrm{dk}%
c(k)\left\vert k\right\rangle
\end{equation}
with $c(k)\overset{L\rightarrow\infty}{\equiv}\sqrt{\frac{L}{2\pi}}c_{n}$ .
The normalization $\left\langle \psi(t)|\psi(t)\right\rangle =1$ implies%
\begin{equation}
1=\left\vert c_{S}\right\vert ^{2}+\sum_{n}\left\vert c_{n}\right\vert
^{2}\overset{L\rightarrow\infty}{\equiv}\left\vert c_{S}\right\vert ^{2}%
+\int_{-\infty}^{+\infty}\mathrm{dk}\left\vert c(k)\right\vert ^{2}\text{ .}%
\end{equation}
In particular, one is typically interested to the case $c_{S}=1$ (but not
only). A formal solution to the time evolution is obtained by introducing the
time-evolution operator:%

\begin{equation}
U(t)=e^{-iHt}\rightarrow\left\vert \psi(t)\right\rangle =U(t)\left\vert
\psi(0)\right\rangle \text{ . }%
\end{equation}
The time-evolution operator $U(t)$ can be expressed in terms of a Fourier
transform (for $t>0$):%
\begin{equation}
U(t)=\frac{i}{2\pi}\int_{-\infty}^{+\infty}\mathrm{dE}\frac{1}%
{E-H+i\varepsilon}e^{-iEt}\text{ }=\frac{i}{2\pi}\int_{-\infty}^{+\infty
}\mathrm{dE}G(E)e^{-iEt}\text{ with }G(E)=\frac{1}{E-H+i\varepsilon}\text{ ,}
\label{uoft}%
\end{equation}
where $\varepsilon$ is an infinitesimal quantity and $G(E)$ is the `propagator
operator', which can be expanded as:%

\begin{equation}
G(E)=\frac{1}{E-H+i\varepsilon}=\sum_{n=0}^{\infty}\left(  \frac{1}%
{E-H_{0}+i\varepsilon}H_{1}\right)  ^{n}\frac{1}{E-H_{0}+i\varepsilon}\text{
,}%
\end{equation}
where we have used that $(AB)^{-1}=B^{-1}A^{-1}$ ($A,B$ arbitrary operators on
the Hilbert space $\mathcal{H}$).

\subsection{Propagator,\ Feynman rules, and survival probability}

We are interested in the evaluation of the survival (or non-decay) probability
amplitude $a_{S}(t)=\left\langle S\right\vert U(t)\left\vert S\right\rangle $,
out of which the survival probability of the state $S$ reads $p_{S}%
(t)=\left\vert a_{S}(t)\right\vert ^{2}.$ In the \textit{trivial limit,} in
which $H=H_{0}$ ( $g\rightarrow0$), one has
\begin{equation}
a_{S}(t)=\left\langle S\right\vert U(t)\left\vert S\right\rangle =\left\langle
S\right\vert e^{-iH_{0}t}\left\vert S\right\rangle =e^{-iM_{0}t}\rightarrow
p_{S}(t)=1\text{ .}%
\end{equation}
Alternatively, one may use Eq. (\ref{uoft}):
\begin{equation}
a_{S}(t)=\left\langle S\right\vert U(t)\left\vert S\right\rangle =\frac
{i}{2\pi}\int_{-\infty}^{+\infty}\mathrm{dE}\frac{1}{E-M_{0}+i\varepsilon
}e^{-iEt}\text{ }=e^{-iM_{0}t}\text{ ,}%
\end{equation}
where we have closed downwards and picked up the pole for $E=M_{0}%
-i\varepsilon$ (one is obliged to close downwards to guarantee convergence).
In passing by, we note that the object
\begin{equation}
G_{S}^{\text{free}}(E)=G_{S}^{(0)}(E)=\left\langle S\right\vert \frac
{1}{E-H_{0}+i\varepsilon}\left\vert S\right\rangle =\frac{1}{E-M_{0}%
+i\varepsilon}%
\end{equation}
is the free propagator of the state $S.$

In the \textit{interacting case }the evaluation of\textit{ }$a(t)$\textit{
}proceeds as follow:
\begin{equation}
a_{S}(t)=\frac{i}{2\pi}\int_{-\infty}^{+\infty}\mathrm{dE}G_{S}(E)e^{-iEt}%
\text{ , where }G_{S}(E)=\left\langle S\right\vert G(E)\left\vert
S\right\rangle =\left\langle S\right\vert \frac{1}{E-H+i\varepsilon}\left\vert
S\right\rangle \label{as1}%
\end{equation}
is the full propagator of $S.$ It is now necessary to evaluate $G_{S}(E)$
explicitly though a lengthy but straightforward calculation \cite{duecan}:
\begin{equation}
G_{S}(E)=\left\langle S\right\vert G(E)\left\vert S\right\rangle =\sum
_{n=0}^{\infty}\left\langle S\right\vert \left(  \frac{1}{E-H_{0}%
+i\varepsilon}H_{1}\right)  ^{n}\frac{1}{E-H_{0}+i\varepsilon}\left\vert
S\right\rangle =\sum_{n=0}^{\infty}G_{S}^{(n)}(E)
\end{equation}
with%
\begin{equation}
G_{S}^{(n)}(E)=\left\langle S\right\vert \left(  \frac{1}{E-H_{0}%
+i\varepsilon}H_{1}\right)  ^{n}\left\vert S\right\rangle \frac{1}%
{E-M_{0}+i\varepsilon}\text{ .}%
\end{equation}
Let us evaluate the first three terms:%
\begin{equation}
n=0\rightarrow G_{S}^{(0)}(E)=\left\langle S\right\vert 1\left\vert
S\right\rangle \frac{1}{E-M_{0}+i\varepsilon}=\frac{1}{E-M_{0}+i\varepsilon
}\text{ ,}%
\end{equation}%
\begin{equation}
n=1\rightarrow\left\langle S\right\vert \frac{1}{E-H_{0}+i\varepsilon}%
H_{1}\left\vert S\right\rangle \frac{1}{E-M_{0}+i\varepsilon}=0\text{ ,}%
\end{equation}

\begin{align}
n  &  =2\rightarrow G_{S}^{(1)}(E)=\left\langle S\right\vert \left(  \frac
{1}{E-H_{0}+i\varepsilon}H_{1}\right)  ^{2}\left\vert S\right\rangle \frac
{1}{E-M_{0}+i\varepsilon}\\
&  =\frac{1}{E-M_{0}+i\varepsilon}\left\langle S\right\vert H_{1}\frac
{1}{E-H_{0}+i\varepsilon}H_{1}\left\vert S\right\rangle \frac{1}%
{E-M_{0}+i\varepsilon}=-\frac{\Pi(E)}{\left(  E-M_{0}+i\varepsilon\right)
^{3}}\text{ .}%
\end{align}
The recursive quantity $\Pi(E)$ reads:%
\begin{equation}
\Pi(E)=-\left\langle S\right\vert H_{1}\frac{1}{E-H_{0}+i\varepsilon}%
H_{1}\left\vert S\right\rangle \text{ .}%
\end{equation}
We introduce $1_{\mathcal{H}}=\left\vert S\right\rangle \left\langle
S\right\vert +\int_{-\infty}^{+\infty}dk\left\vert k\right\rangle \left\langle
k\right\vert $ two times, obtaining:
\begin{align}
\Pi(E)  &  =-\left\langle S\right\vert H_{1}1_{\mathcal{H}}\frac{1}%
{E-H_{0}+i\varepsilon}1_{\mathcal{H}}H_{1}\left\vert S\right\rangle
=-\int_{-\infty}^{+\infty}\mathrm{dk}\int_{-\infty}^{+\infty}\mathrm{dq}%
\left\langle S\right\vert H_{1}\left\vert k\right\rangle \left\langle
k\right\vert \frac{1}{E-H_{0}+i\varepsilon}\left\vert q\right\rangle
\left\langle q\right\vert H_{1}\left\vert S\right\rangle \nonumber\\
&  =-\int_{-\infty}^{+\infty}\mathrm{dk}\int_{-\infty}^{+\infty}%
\mathrm{dq}\frac{gf(k)}{\sqrt{2\pi}}\frac{\delta(k-q)}{E-\omega
(k)+i\varepsilon}\frac{gf(q)}{\sqrt{2\pi}}=-\int_{-\infty}^{+\infty}%
\frac{\mathrm{dk}}{2\pi}\frac{g^{2}f(k)^{2}}{E-\omega(k)+i\varepsilon}\text{
,}\nonumber
\end{align}
where $\left\langle S\right\vert H_{1}\left\vert k\right\rangle =gf(k)/\sqrt
{2\pi}$ was used. Going further, for $n=0,1,2,...$ we get $G_{S}%
^{(2n+1)}(E)=0$ and
\begin{equation}
G_{S}^{(2n)}(E)=\frac{\left[  -\Pi(E)\right]  ^{n}}{\left(  E-M_{0}%
+i\varepsilon\right)  ^{n+1}}\text{ .}%
\end{equation}
Finally:%
\begin{align}
G_{S}(E)  &  =\sum_{n=0}^{\infty}G_{S}^{(2n)}(E)=\sum_{n=0}^{\infty}%
\frac{\left[  -\Pi(E)\right]  ^{n}}{\left(  E-M_{0}+i\varepsilon\right)
^{2n+1}}=\frac{1}{\left(  E-M_{0}+i\varepsilon\right)  }\sum_{n=0}^{\infty
}\frac{\left[  -\Pi(E)\right]  ^{n}}{\left(  E-M_{0}+i\varepsilon\right)
^{n}}\\
&  =\frac{1}{\left(  E-M_{0}+i\varepsilon\right)  }\frac{1}{1+\frac{\Pi
(E)}{E-M_{0}+i\varepsilon}}=\frac{1}{E-M_{0}+\Pi(E)+i\varepsilon}\text{ .}%
\end{align}
At this point, we can identify `Feynman rules' reminiscent of QFT:%
\begin{align}
\text{bare }S\text{ propagator }  &  \rightarrow\frac{1}{E-M_{0}+i\varepsilon
}\\
\text{bare }k\text{ propagator (}k\text{ fixed)}  &  \rightarrow\frac
{1}{E-\omega(k)+i\varepsilon}\\
kS\text{ vertex}  &  \rightarrow gf(k)\\
\text{internal }k\text{ line(}k\text{ not fixed)}  &  \rightarrow-\Pi
(E)=\int_{-\infty}^{+\infty}\frac{dk}{2\pi}\frac{g^{2}f(k)^{2}}{E-\omega
(k)+i\varepsilon}%
\end{align}
Note, the latter can be understood by applying $gf(k)$ at each vertex and the
$k$-propagator in the middle, and by an overall integration $\int_{-\infty
}^{+\infty}\frac{dk}{2\pi}$ due to the fact that $k$ is not fixed.

The full propagator of $S$ determined above,%
\begin{equation}
\text{full }S\text{ propagator }\rightarrow\frac{1}{E-M_{0}+\Pi
(E)+i\varepsilon}\text{ ,}%
\end{equation}

can be also obtained in a very elegant way by using the Bethe-Salpeter
equation obtained by using the Feynman rules listed above:
\begin{equation}
G_{S}(E)=\frac{1}{E-M_{0}+i\varepsilon}-\frac{1}{E-M_{0}+i\varepsilon}%
\Pi(E)G_{S}(E)\text{ .}%
\end{equation}

Finally, the survival amplitude (\ref{as1}) can be expressed as:%
\begin{equation}
a_{S}(t)=\frac{i}{2\pi}\int_{-\infty}^{+\infty}\mathrm{dE}G_{S}(E)e^{-iEt}%
=\frac{i}{2\pi}\int_{-\infty}^{+\infty}\mathrm{dE}\frac{1}{E-M_{0}%
+\Pi(E)+i\varepsilon}e^{-iEt}\text{ .} \label{as2}%
\end{equation}

\subsection{Spectral function and survival probability}

Let us denote the basis of eigenstates of the Hamiltonian $H$ as $\left\vert
m\right\rangle $ with%
\begin{equation}
H\left\vert m\right\rangle =m\left\vert m\right\rangle \text{ for }m\geq
m_{th}\text{ (}m_{th}\text{ is the low-energy threshold) .}%
\end{equation}
The existence of a minimal energy $m_{th}$ is a general physical and
mathematical property. The states $\left\vert m\right\rangle $ form an
orthonormal basis of the Hilbert space $\mathcal{H=}\{\left\vert
m\right\rangle $ with $m\geq m_{th}\}$, whose elements fulfill standard
relations:
\begin{equation}
1_{\mathcal{H}}=\int_{m_{th}}^{+\infty}\mathrm{dm}\left\vert m\right\rangle
\left\langle m\right\vert \text{ ; }\left\langle m_{1}|m_{2}\right\rangle
=\delta(m_{1}-m_{2})\text{ .}%
\end{equation}
The link between the `old' basis $\{\left\vert S\right\rangle ,\left\vert
k\right\rangle \}$ (eigenstates of $H_{0}$) and the `new' basis $\{\left\vert
m\right\rangle \}$ (eigenstates of $H$) is not trivial. The state $\left\vert
S\right\rangle $ can be expressed in terms of the basis $\{\left\vert
m\right\rangle $ $\}$ as
\begin{equation}
\left\vert S\right\rangle =\int_{m_{th}}^{\infty}\mathrm{dm}\alpha
_{S}(m)\left\vert m\right\rangle \text{ with }\alpha_{S}(m)=\left\langle
S|m\right\rangle \text{ .}%
\end{equation}
The quantity
\begin{equation}
d_{S}(m)=\left\vert \alpha_{S}(m)\right\vert ^{2}=\left\vert \left\langle
S|m\right\rangle \right\vert ^{2}%
\end{equation}
is called the \textbf{spectral function (or energy/mass distribution) }of the
state $S$. The normalization of the state $\left\vert S\right\rangle $ implies
the normalization of the mass distribution $d_{S}(m)$:
\begin{equation}
1=\left\langle S|S\right\rangle =\int_{m_{th}}^{\infty}d_{S}(m)\mathrm{dm}%
\text{ .} \label{norm}%
\end{equation}
The simple intuitive interpretation is that $d_{S}(m)\mathrm{dm}$ represents
the probability that the state $S$ has a energy (or mass) between $m$ and
$m+dm.$ As a consequence, the time-evolution can be easily evaluated by
inserting $1=\int_{m_{th}}^{+\infty}\mathrm{dm}\left\vert m\right\rangle
\left\langle m\right\vert $ two times:%
\begin{equation}
a_{S}(t)=\left\langle S\right\vert U(t)\left\vert S\right\rangle =\left\langle
S\right\vert e^{-iHt}\left\vert S\right\rangle =\int_{m_{th}}^{\infty
}\mathrm{dm}d_{S}(m)e^{-imt}\text{ .}%
\end{equation}
This is all formally correct, but it does not help us further as long as we do
not have a way to calculate $d_{S}(m)$. This is possible by using the
propagator of $S$ studied in Sec. 3.2. In fact, the propagator can be
re-expressed as (again, inserting $1=\int_{m_{th}}^{+\infty}dm\left\vert
m\right\rangle \left\langle m\right\vert $ two times):%
\begin{equation}
G_{S}(E)=\frac{1}{E-M_{0}+\Pi(E)+i\varepsilon}=\left\langle S\right\vert
1\frac{1}{E-H+i\varepsilon}1\left\vert S\right\rangle =\int_{m_{th}}^{+\infty
}\mathrm{dm}\frac{d_{S}(m)}{E-m+i\varepsilon}\text{ .} \label{kl}%
\end{equation}
Its physical meaning can be understood by noticing that the dressed propagator
$G_{S}(E)$ has been rewritten as the `sum' of free propagators, whose weight
function is $d_{S}(m).$ As a next step, we need to invert Eq. (\ref{kl}). Let
us first consider the case $g=0.$ In this limit, it is evident from Eq.
(\ref{kl}) that:
\begin{equation}
d_{S}(E)=\delta(E-M_{0})\text{ .}%
\end{equation}
This is expected because in this case the state $\left\vert S\right\rangle $
is an eigenstate of the Hamiltonian, hence the mass distribution is a
delta-function peaked at $M_{0}$. When the interaction is switched on, we
evaluate the imaginary part of Eq. (\ref{kl}):
\begin{equation}
\operatorname{Im}G_{S}(E)=\int_{m_{th}}^{+\infty}\mathrm{dm}\frac{-\varepsilon
d_{S}(m)}{(E-m)^{2}+\varepsilon^{2}}=-\int_{m_{th}}^{+\infty}\mathrm{dm}%
d_{S}(m)\pi\delta(E-m)=-\pi d_{S}(E)\text{ }.
\end{equation}
Hence $d_{S}(E)$ is calculated as:%
\begin{equation}
d_{S}(E)=-\frac{\operatorname{Im}G_{S}(E)}{\pi}=\frac{1}{\pi}\frac
{\operatorname{Im}\Pi(E)}{(E-M_{0}+\operatorname{Re}\Pi(E))^{2}+\left(
\operatorname{Im}\Pi(E)\right)  ^{2}}\text{ .} \label{ds}%
\end{equation}
The normalization of $d_{S}(E),$ Eq. (\ref{norm}), can be also proven by using
Eq. (\ref{ds}), see details in\ Ref. \cite{gpprd}.

In the end, once the spectral function $d_{S}(m)$ is known, the survival
amplitude can be re-expressed as its Fourier transform by using Eqs.
(\ref{as2}) and (\ref{kl}):%
\begin{align}
a_{S}(t)  &  =\frac{i}{2\pi}\int_{-\infty}^{+\infty}\mathrm{dE}G_{S}%
(E)e^{-iEt}=\frac{i}{2\pi}\int_{-\infty}^{+\infty}\mathrm{d}E\int_{m_{th}%
}^{+\infty}\mathrm{dm}\frac{d_{S}(m)}{E-m+i\varepsilon}e^{-imt}\nonumber\\
&  =\int_{m_{th}}^{+\infty}\mathrm{dm}d_{S}(m)e^{-imt}=\int_{m_{th}}^{+\infty
}\mathrm{dm}d_{S}(m)e^{-imt}\text{ .}%
\end{align}
The latter expression coincides with Eq. (1), whose detailed determination was
our goal. From here on, all the usual strategy can be applied
\cite{ghirardi,fprev,koshinorev,duecan}. In particular, the (unphysical)
Breit-Wigner limit is obtained for $\omega(k)=k$ (unlimited from below) and
$f(k)=1,$ out of which $d_{S}(m)=\frac{\Gamma}{2\pi}\left(  \left(
m-M_{0}\right)  ^{2}+\Gamma^{2}/4\right)  ^{-1}$ with $\Gamma=g^{2}.$ In this
case, $a_{S}(t)=e^{-i(M_{0}-i\Gamma/2)t}$ and $p_{S}(t)=e^{-\Gamma t}$ (see
details in\ Refs. \cite{giacosapra,actalast}).

\section{Conclusions}

We have proven that Eq. (1) holds in the QFT-like approach of effective Lee
Hamiltonians by showing all the main steps leading to it. However, a Lee
Hamiltonian is not fully equivalent to QFT, since some features are still
missing. In fact, the Lee approach does not contain transitions from the
vacuum state to some particles (in genuine QFT, terms of the type $\left\vert
0\right\rangle \left\langle S\varphi_{1}\varphi_{2}\right\vert $+hc are also
part of the interacting Hamiltonian and affect the results for finite time
intervals). Moreover, quadratic expressions are not present in the
propagator(s) of the Lee Hamiltonian but naturally appear in QFT.

Hence, the main question for the future reads: is Eq. (1) as it stands valid
also in\ QFT? If, as argued in\ Refs. \cite{duecan,zenoqft}, this is true,
non-exponential decay is realized in QFT both at short and long times.
Moreover, in the interesting case of a super-renormalizable Lagrangian, the
short-time behavior is independent on the cutoff (this is so because the
energy distribution $d_{S}(m)$ scales as $m^{-3}$ for large $m$). Note, this
is different from the result of Ref. \cite{maiani} obtained by using
perturbation theory at second order (that is, without resumming the propagator).

A further interesting topic for the future is the study of the decay of a
particle with a nonzero momentum
\cite{khalfin2,stefanovich,shirokov,urbanowski,giunti,rel}. Contrary to naive
expectations, the usual relativistic time dilatation formula does not hold
(even in the exponential limit, a different analytical result is obtained, see
details in Ref. \cite{rel}). The full understanding of the study of decay in
QFT can also help to shed light on decays of moving particles.

\bigskip

\textbf{Acknowledgments: }F. G. thanks S. Mr\'{o}wczy\'{n}ski and G. Pagliara
for very useful discussions. F. G. acknowledges support from the Polish
National Science Centre (NCN) through the OPUS project no. 2015/17/B/ST2/01625.

\end{document}